# Hformer: Hybrid CNN-Transformer for Fringe Order Prediction in Phase Unwrapping of Fringe Projection

XINJUN ZHU,[1,*] ZHIQIANG HAN,[1] MENGKAI YUAN,[1] QINGHUA GUO,[2] AND HONGYI WANG[1]

[1]*School of Artificial Intelligence, Tiangong University, Tianjin 300387, China*
[2]*School of Electrical, Computer and Tele communications Engineering, University of Wollongong, Wollongong, NSW 2522, Australia*
*\* xinjunzhu@tiangong.edu.cn*

**Abstract:** Recently, deep learning has attracted more and more attention in phase unwrapping of fringe projection three-dimensional (3D) measurement, with the aim to improve the performance leveraging the powerful Convolutional Neural Network (CNN) models. In this paper, for the first time (to the best of our knowledge), we introduce the Transformer into the phase unwrapping which is different from CNN and propose Hformer model dedicated to phase unwrapping via fringe order prediction. The proposed model has a hybrid CNN-Transformer architecture that is mainly composed of backbone, encoder and decoder to take advantage of both CNN and Transformer. Encoder and decoder with cross attention are designed for the fringe order prediction. Experimental results show that the proposed Hformer model achieves better performance in fringe order prediction compared with the CNN models such as U-Net and DCNN. Moreover, ablation study on Hformer is made to verify the improved feature pyramid networks (FPN) and testing strategy with flipping in the predicted fringe order. Our work opens an alternative way to deep learning based phase unwrapping methods, which are dominated by CNN in fringe projection 3D measurement.

## 1. Introduction

In fringe projection three-dimensional (3D) measurement, phase unwrapping is critical since extracted phase through phase retrieval methods such as the phase shifting method and the Fourier transform method is wrapped within the range from $-\pi$ to $\pi$, leading to $2\pi$ discontinuities due to the arctangent operator [1-6]. The extracted phase is called wrapped phase. Hence, the continuous (absolute or unwrapped) phase related to the 3D information of the measured objects should be recovered from the wrapped phase through phase unwrapping [1-3]. Despite recent advances on phase unwrapping, robust and accurate phase unwrapping still remains as a challenging issue, due to the existence of noise, abrupt changes in the objects, and other factors [2]. To tackle the phase unwrapping problem, numerous methods have been proposed, which can be categorized into spatial phase unwrapping [7-9] and temporal phase unwrapping [3,10] methods as well as recently proposed deep learning based methods [11-14].

Essentially, phase unwrapping is to determine the fringe order for each point in the wrapped phase, which can be formulated as a classification problem and addressed by deep learning. By now, deep learning based phase unwrapping methods have attracted more and more attention through end-to-end learning from wrapped phase to fringe order. Inspired by the great success of convolutional neural network (CNN) in many fields [15-18], CNN is also used to learn the mapping between the input wrapped phase and output fringe order in phase unwrapping. For instance, Spoorthi *et al*. [11] firstly proposed PhaseNet to phase unwrapping. In the method, the phase unwrapping problem is formulated as a dense classification problem and a U-Net based neural network is trained to predict the fringe order at each pixel from the wrapped phase maps. Subsequently the improved version (PhaseNet 2.0) [12] was proposed based on DenseNet to improve the accuracy of phase unwrapping. Qian *et al*. [13] introduced deep learning in the phase unwrapping of binocular stereo fringe projection 3D measurement where the left and right fringe patterns are used to produce the geometric constraints in CNN training and testing. In addition, Yin *et al*. [14] introduced deep learning to more frame wrapped phases in temporal based deep learning phase unwrapping. In their method, the mapping between the input temporal wrapped phases with two frequencies and the output fringe order is learned using CNN.

Although CNN has achieved good performance in phase unwrapping in the above works, it is lack of global receptive field in theory, which is difficult to model the global dependency of the image [19]. In contrast to CNN, the Transformer, which has a global receptive field, has shown great performance in the field of Natural Language Processing (NLP) and Computer Vision (CV) recently. The Transformer is an architecture that uses self-attention mechanism as the basic layer to extract features in sequence data or images. Starting with the ViT [20], more and more Transformer based models have been used in various downstream tasks including classification, segmentation, detection, image restoration, achieving the state-of-the-art performance. For instance, the Swin-Transformer [21] achieves the highest accuracy in image recognition, and the Uformer [22] excels in the field of image restoration and image denoising.

To the best of our knowledge, there has been no work reported for the development of Transformer based phase unwrapping especially in fringe projection. In this work, for the first time, we explore the advantages of the Transformer in phase unwrapping, and demonstrate the improved phase unwrapping results over CNN. This work opens an alternative way to achieve phase unwrapping. The main contributions of our work are as follows:

(1) For the first time, we introduce Transformer based on the self-attention mechanism into fringe projection, and explore its application to phase unwrapping via fringe order prediction.

(2) We propose a hybrid CNN-Transformer model called Hformer to predict fringe order. The developed model mainly includes backbone, encoder and decoder, which is able to combine the advantages of CNN and Transformer. Encoder and decoder with cross attention that belong to Transformer are designed for the phase unwrapping. Experimental results show the superiority over existing CNN phase unwrapping methods.

(3) An improved feature pyramid network (FPN) is proposed in Hformer to merge multi-scale features to improve the accuracy of fringe order prediction. The ablation study proves that the improved FPN and testing strategy with flipping can improve the accuracy of fringe order prediction.

The organization of this paper is as follows: Section 2 presents our method of phase unwrapping with the proposed Hformer model. The architecture of the proposed model devoted to training and prediction for fringe order is developed. The encoder and decoder components for the Transformer are designed with a Cross Attention Transformer (CAT) block. An improved FPN is proposed to fuse multi-scale feature maps from the CAT block in the decoder. Section 3 provides the experimental results, including the performance comparisons with CNN based methods as well ablation study on the proposed model, followed by conclusions drawn in Section 4.

## 2. Proposed method to phase unwrapping

### 2.1 Architecture overview of the proposed Hformer model

The Hformer for fringe order prediction we propose in this work is composed of a backbone, encoder and decoder, where the backbone corresponds to the CNN, and the encoder and decoder correspond to the Transformer. HRNet [23] is selected as the backbone for the advantages of parallel and high resolution. Cross Attention Transformer (CAT) block [24] is used as the basic unit of the encoder and decoder for the advantages of alternating attention inner the image patch instead of the whole image to capture local information and apply attention between image patches to capture global information. In order to fuse multi-scale feature maps, skip connections between the encoder and decoder is employed and the improved FPN is designed to fuse the multi-scale feature maps of the decoder. The overall architecture of Hformer is shown in Figure 1.

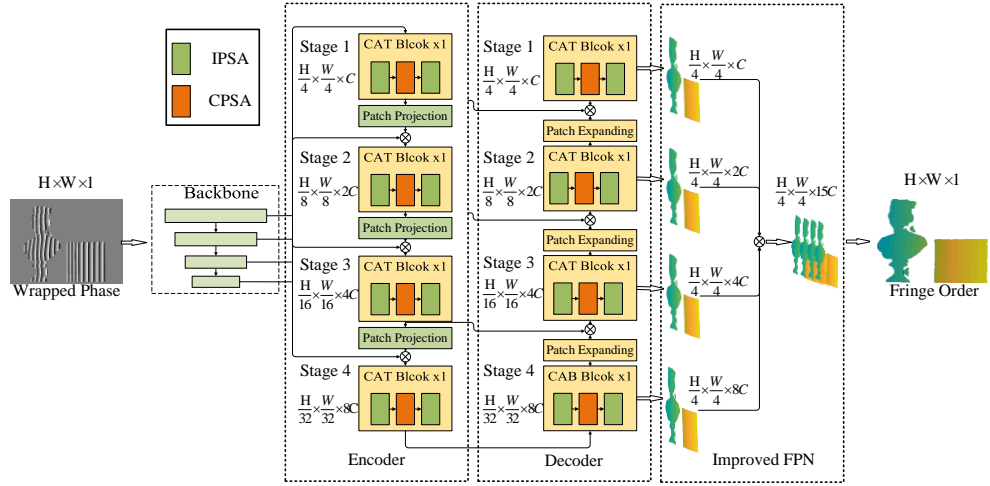

Fig. 1. The architecture of Hformer.

Firstly, as shown in Fig. 1, four scale feature maps are obtained after passing the wrapped phase with a size of H×W×1 through the backbone. Then, the feature maps of different scales and absolute position information are provided to the CAT block of encoder. Specifically, the first feature maps are input to the first stage of the encoder, which are H/4×W/4×C (C is the number of first feature maps), and then they are down-sampled through the Patch Projection [21,24] before entering the second stage, with the size of the feature maps decreased by 1/2 times and the number increased by 4 times. Except the fourth stage of the encoder, each stage consists of a CAT block and a Patch Projection.

Secondly, inspired by U-Net [25], we designed a symmetric Transformer decoder to the encoder. The decoder is composed of a CAT block and a Patch Expanding [26]. It reduces the number of the feature maps to 1/4 and increases the size of the feature maps by 2 times. Context features extracted by the encoder are connected to encoder through skip connections to recover spatial information caused by down-sampling. Similar to encoder, except the first stage of the decoder, each stage of the decoder consists of a CAT block and a Patch Expanding.

Finally, unlike the U-shaped architecture which produces output feature maps only in the last stage, the proposed Hformer model uses an improved FPN to fuse the output feature maps of different stages of decoder. Specifically, the feature maps of the four stages of the decoder are interpolated by upsampling to the size of the first stage of the decoder to concatenate together, and to output the pixel-level prediction through a convolutional layer. The final pixel-level prediction output results are the fringe orders needed in phase unwrapping during training, validation and testing.

*2.2 Cross attention transformer block*

CAT block is a kind of attention mechanism, which alternates attention within image patches instead of the whole image to capture local information, and applies attention between image blocks divided from single-channel feature maps to capture global information. Figure 2 shows the components of the CAT block, which consists of Layer norm (LN), Inner-patch self-attention (IPSA), Cross-patch self-attention (CPSA), Skip connections and activated 3-layer

Multilayer perceptron (MLP) with Linear unit with gaussian error (GELU). A CAT block [24] can be formulated as:

$$\mathbf{I}_1^p = \text{IPSA}\left(\text{LN}\left(\mathbf{x}^p\right)\right) + \mathbf{x}^p \tag{1}$$

$$\mathbf{M}_1^p = \text{MLP}\left(\text{LN}\left(\mathbf{I}_1^p\right)\right) + \mathbf{I}_1^p \tag{2}$$

$$\mathbf{C}^p = \text{CPSA}\left(\text{LN}\left(\mathbf{M}_1^p\right)\right) + \mathbf{M}_1^p \tag{3}$$

$$\mathbf{M}_2^p = \text{MLP}\left(\text{LN}\left(\mathbf{C}^p\right)\right) + \mathbf{C}^p \tag{4}$$

$$\mathbf{I}_2^p = \text{IPSA}\left(\text{LN}\left(\mathbf{M}_2^p\right)\right) + \mathbf{M}_2^p \tag{5}$$

$$\mathbf{y}^p = \text{MLP}\left(\text{LN}\left(\mathbf{I}_2^p\right)\right) + \mathbf{I}_2^p \tag{6}$$

where $\mathbf{x}^p$ and $\mathbf{y}^p$ denotes input feature maps and output feature maps of the $p^{th}$ block respectively, $\mathbf{C}^p$, $\mathbf{I}_u^p (u=1,2)$ and $\mathbf{M}_u^p (u=1,2)$ denote the outputs of the CPSA module, IPSA module and MLP module of the $p^{th}$ block, respectively. IPSA calculates the attention matrix within blocks while CPSA calculates the attention matrix between blocks. IPSA and CPSA have the same self-attention calculation as in [21], which can be expressed as follows:

$$\text{Attention}(\mathbf{Q},\mathbf{K},\mathbf{V}) = \text{SoftMax}\left(\frac{\mathbf{Q}\mathbf{K}^T}{\sqrt{d}} + \mathbf{B}\right)\mathbf{V} \tag{7}$$

where $\mathbf{Q}$, $\mathbf{K}$, and $\mathbf{V}$ represent the query, key and value matrixes respectively, and $\mathbf{B}$ is the relative position deviation, $d$ is the query and key matrixes dimension, $T$ denotes transposition of matrix, and SoftMax function is used in the multi-classification, which maps the output of multiple neurons to the (0,1) interval.

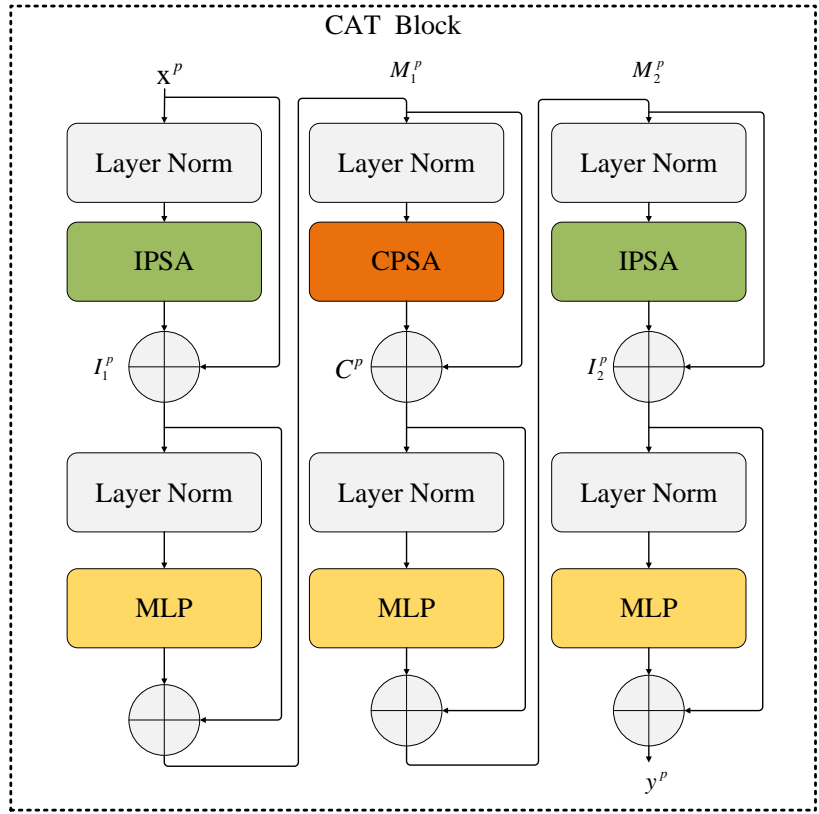

Fig. 2. Architecture of CAT block in Hformer.

*2.3 Improved FPN*

The output multi-scale feature maps from the CAT block in the decoder are fused in the FPN. However, the original FPN [27] needs a huge memory, which gives rise to the need of a huge memory of the Hformer, making it difficult to complete end-to-end training. Moreover, the other stages output feature maps are different from that of the last stage in terms of output size for the original FPN, which are not available for the task of fringe order prediction. To tackle these problems, we proposed an improved FPN as shown in Figure 3, where the blue layers represent the input feature maps while the yellow layers represent output feature maps. As shown in Fig. 3, the improved FPN (Fig. 3(c)) still maintains a single output form (Fig. 3(a)), but at the same time integrates feature maps of different scales from other three stages, compared with the original FPN (Fig. 3(b)) that requires four output feature maps. The performance of improved FPN will be discussed in detail in Section 3.4.

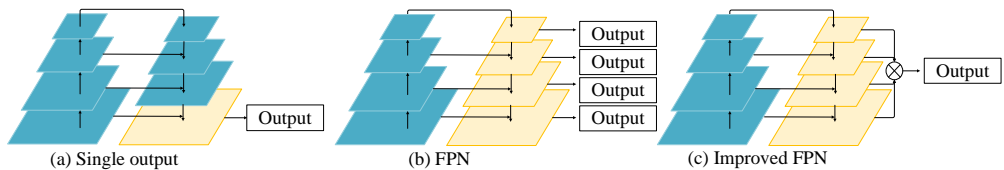

Fig. 3. Different fusion output methods. (a), (b) and (c) respectively represent single output, (original) FPN output and improved FPN output.

## 3. Experiments

### 3.1 Dataset

This dataset used in this work is from Ref. [13], which includes 1000 wrapped phase and fringe order images with a resolution of 640×480 pixels. The wrapped phase can be obtained with the phase shifting method as follows:

$$\varphi(x, y) = \arctan \frac{\sum_{n=0}^{N-1} I_n(x, y) \sin\left(\frac{2\pi n}{N}\right)}{\sum_{n=0}^{N-1} I_n(x, y) \cos\left(\frac{2\pi n}{N}\right)} \tag{8}$$

where $\varphi(x, y)$ is wrapped phase, and $I_n(x, y)$ is the $(n+1)$-th captured fringe pattern image, $n = 0, 1, ..., N-1$.

Mathematically, the relationship between the wrapped phase $\varphi(x, y)$ and unwrapped phase $\Phi(x, y)$ can be described as:

$$\Phi(x, y) = \varphi(x, y) + k(x, y) \times 2\pi \tag{9}$$

where $k(x, y)$ is the fringe order.

As to the dataset, the wrapped phase samples are obtained using the three-step phase shift algorithm with $N=3$ in Eq. (8) and the fringe order samples are obtained by using triple-camera stereo phase unwrapping and adaptive depth constraint methods described in [13], which serve the ground truth. The range of label (fringe order) is 0-33. In the experiment, 856 samples are used for training, 72 samples are used for validation, 72 samples are used for test. During the training, an image is divided into 4 partially overlapped images with image size of 384×384, amounting to 3424 in total. As to the validation and testing, 72 validation and 72 testing images are used with a resolution of 480×640. In detail, sliding window with window size 384×384 is used for verification and test to output predicted results from the overlapped patches of validation and testing images and they are synthesized into a whole image. The sliding step for window is set as 256 and 96 in the width and height directions of the images. In this paper, Mean Intersection over Union (mIoU), Mean Square Error (MSE) and Mean Absolute Error (MAE) are used as the criteria for performance evaluation.

### 3.2 Implementation details

All models are implemented with Python 3.7 and PaddlePaddle 2.0 under same hyper-parameter setting on the Nvidia V100 GPU with 32GB memory. Cross-entropy is employed as loss function. Both horizontal and vertical flips are used for data augmentation during training and testing. The batch size is 4 in the training stage. Stochastic gradient descent (SGD) optimizer is used, where momentum is 0.9 and weight decay is 0.01. The initial learning rate is 0.0005, and the decay strategy is polynomial learning rate policy with power 0.9. The number of training times is set to 40,000 in order to choose the best model for the three methods.

### 3.3 Experimental results

Table 1 shows the performances of proposed model as well as U-Net and DCNN to the fringe order prediction in terms of mIoU, MSE and MAE. It is demonstrated that the proposed Hformer model achieves the best performance among the three models, with mIoU, MSE and MAE of 0.6754, 0.5047 and 0.0574 respectively. In contrast, the mIoU, MSE and MAE for U-Net is 0.2030, 1.9985 and 0.2879. According to [12], a DCNN model based on DenseNet block was built for comparison [28]. As to the DCNN, the corresponding indexes are 0.2677, 7.7575, 0.7533, respectively. In summary, the proposed Hformer performs better than U-Net, DCNN in terms of mIoU, MSE and MAE indexes.

Figure 4 shows the predicted fringe order of the above three models for 4 kinds of tested objects. Figs. 4(a), 4(b),4(c) respectively show the predicted fringe order by U-Net, DCNN and Hformer, and Fig. 4 (d) shows the ground truth for comparison. Figure 5 shows the enlarged parts of Fig. 4 with red boxes of different models. As can be seen from Figs. 4 and 5, the prediction of fringe order of U-Net and DCNN are generally different from ground truth, and the fringe order is neither continuous nor smooth in some areas. Erroneous areas of the fringe order for U-Net and DCNN can be found in Fig. 4. In contrast, the fringe order predicted by Hformer is smooth and closer to the ground truth. In conclusion, the proposed Hformer model achieves better visual quality of the predicted results than the other two models.

**Table 1. The mIoU, MSE and MAE of different models on the wrapped phase dataset.**

| Model | mIoU↑ | MSE↓ | MAE↓ |
|---|---|---|---|
| U-Net | 0.2030 | 1.9985 | 0.2879 |
| DCNN | 0.2677 | 7.7575 | 0.7533 |
| Hformer | **0.6754** | **0.5047** | **0.0574** |

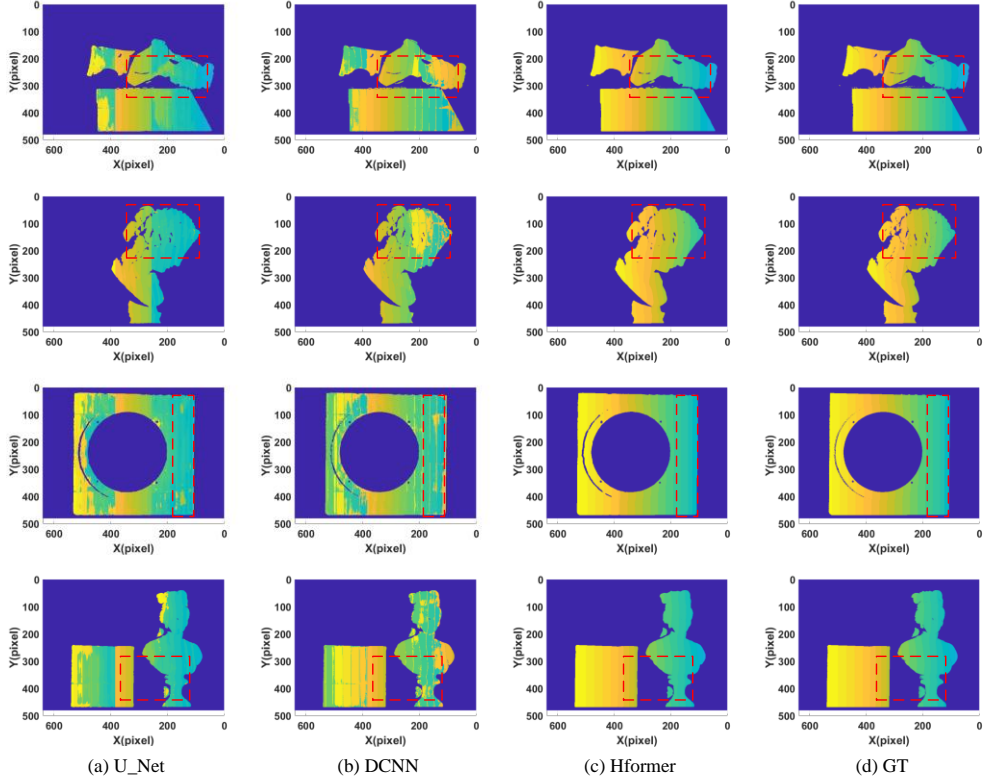

(a) U_Net     (b) DCNN     (c) Hformer     (d) GT

Fig. 4. The fringe order results prediction of wrapped phase. (a), (b) and (c) respectively show predicted fringe order for U-Net, DCNN and Hformer, (d) shows the ground truth of fringe order.

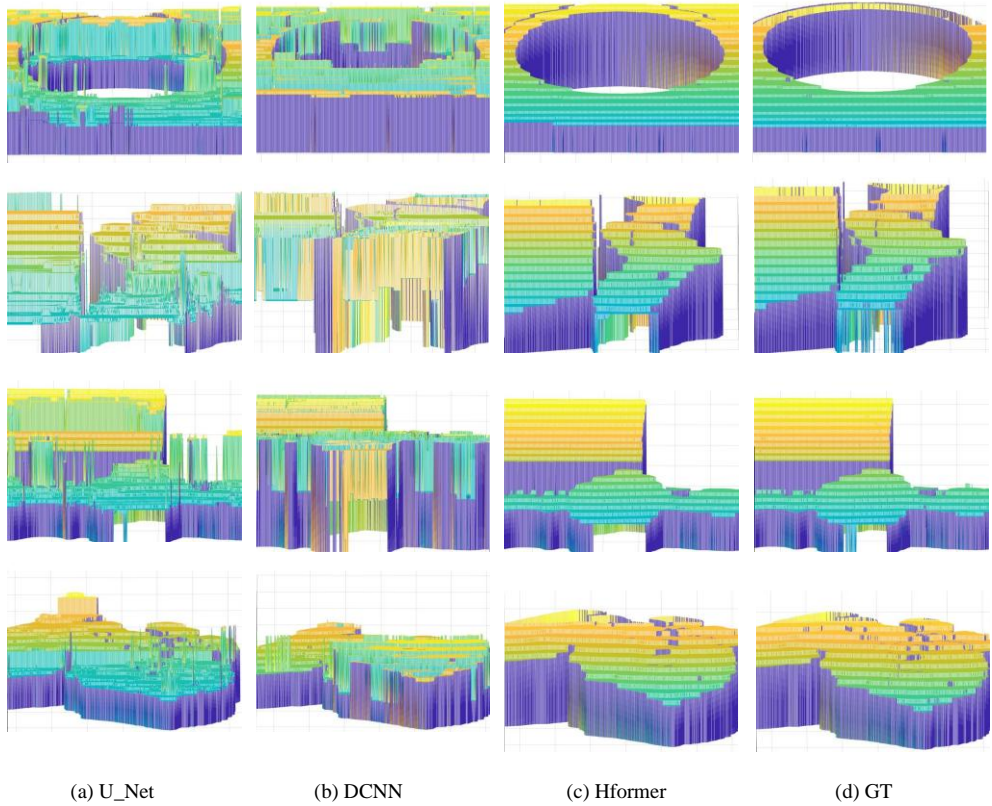

|(a) U_Net|(b) DCNN|(c) Hformer|(d) GT|

Fig. 5. Enlarged parts of prediction results in red box of Figure 4. (a), (b), (c), and (d) respectively show enlarged parts of Figs. 4(a), 4(b), 4(c), and 4(d) with red box.

## 3.4 Ablation study on the proposed model to phase unwrapping

In this section, the ablation experiments are conducted for the architecture on the above dataset, to further investigate the architecture and prediction strategy. In particular, the effects of the improved FPN, number of CAT block, dimensions of CAT block and backbone, horizontal flip and vertical flip on the proposed Hformer model are investigated in this work. Table 2 lists the ablation study results with 8 different conditions including Hformer model without improved FPN, with improved FPN, with more CAT blocks, with larger CAT block, with larger backbone, with horizontal flip, with vertical flip, and lastly with both horizontal flip and vertical flip.

Firstly, Figure 6 shows the prediction fringe order results by Hformer with and without improved FPN. It can be seen from Fig. 6 that the fringe order predicted by the model without improved FPN has a wide range area of errors, while the improved FPN can largely reduce the errors. Table 2 shows that the performance of model with improved FPN has better performance in terms of mIoU, MSE and MAE.

Secondly, Figure 7 shows the predicted fringe order results by the proposed Hformer with the original number and increased number of CAT blocks of encoder and decoder, respectively. Figure 8 shows the predicted fringe order results similar to Fig. 7 instead of the dimensions of CAT block. Figure 9 shows with original and increased dimensions of backbone respectively. It can be seen from Figs. 7,8 and 9 and Table 2 that the increase of the number and the dimension of CAT blocks, and the increase of the dimension of backbone brings worse prediction results indexes, which also lead to high computational cost and time cost.

Lastly, Figure 10 shows the 3 strategies of data augmentation using horizontal flip alone,

vertical flip alone as well as using both horizontal flip and vertical flip in the test. It can be seen from Fig. 10 and Table 2 that the use of both horizontal flip and vertical flip achieves best performance in the three strategies.

Table 2. Performance comparison of models under different ablation study conditions.

| Model | Hformer | | | | | | mIoU↑ | MSE↓ | MAE↓ |
|---|---|---|---|---|---|---|---|---|---|
| | IFPN | DM | LM | LB | HF | VF | | | |
| A | | | | | | | 0.4980 | 1.5014 | 0.1700 |
| B | √ | | | | | | 0.6703 | 0.5695 | 0.0621 |
| C | √ | √ | | | | | 0.6235 | 0.6623 | 0.0794 |
| D | √ | | √ | | | | 0.6179 | 0.5944 | 0.0775 |
| E | √ | | | √ | | | 0.6069 | 0.6205 | 0.0861 |
| F | √ | | | | √ | | 0.6743 | 0.5187 | 0.0575 |
| G | √ | | | | | √ | 0.6711 | 0.5388 | 0.0607 |
| H | √ | | | | √ | √ | **0.6754** | **0.5047** | **0.0574** |

IFPN: the improved FPN, DM: the deeper model (means a greater number of CAT blocks), LM: the larger model (means larger dimension of CAT blocks), LB: the larger backbone (means larger dimension of backbone), HF: the horizontal flip, VF: the vertical flip. A, B, C, D, E, F, G, H respectively denote the base Hformer, with IFPN, with IFPN and DM, with IFPN and LM, with IFPN and LB, with IFPN and HF, with IFPN and VF, and the last with IFPN, HF and VF in Hformer.

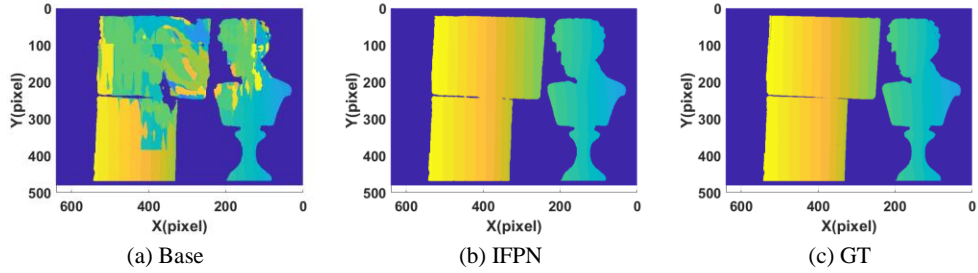

(a) Base      (b) IFPN      (c) GT

Fig. 6. Effect of IFPN on Hformer to fringe order prediction. (a), (b) and (c) respectively show the predicted fringe order by Hformer with Base, with IFPN, and the ground truth of fringe order.

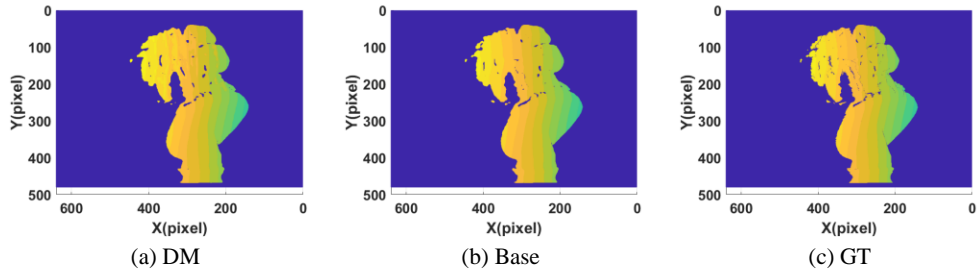

(a) DM      (b) Base      (c) GT

Fig. 7. Effect of DM (the deeper model) on Hformer to fringe order prediction. (a), (b) and (c) respectively show the predicted fringe order by Hformer with DM, with Base, and the ground truth of fringe order.

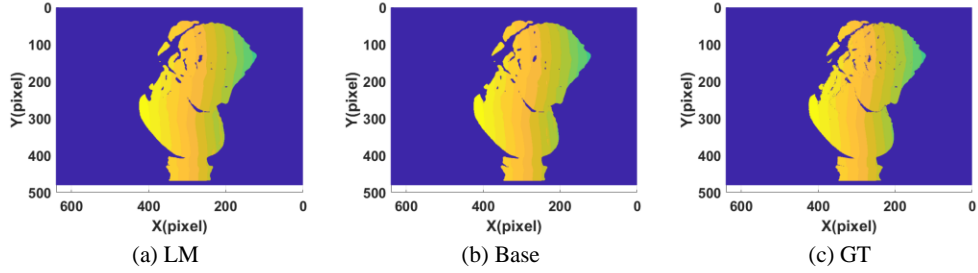

(a) LM    (b) Base    (c) GT

Fig. 8. Effect of LM (the larger model) on Hformer to fringe order prediction. (a), (b) and (c) respectively show the predicted fringe order by Hformer with LM, with Base, and the ground truth of fringe order.

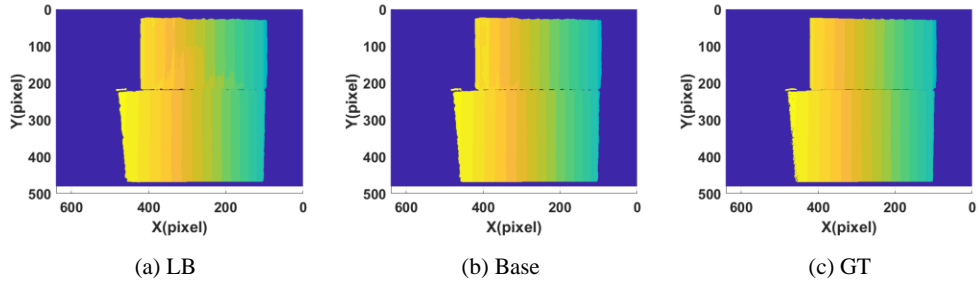

(a) LB    (b) Base    (c) GT

Fig. 9. Effect of LB (the larger backbone) on Hformer to fringe order prediction. (a), (b) and (c) respectively show the predicted fringe order by Hformer with LB, with Base, and the ground truth of fringe order.

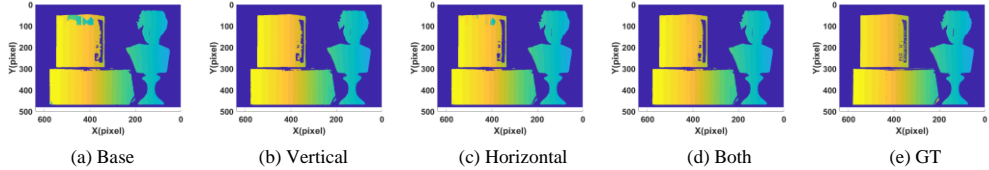

(a) Base    (b) Vertical    (c) Horizontal    (d) Both    (e) GT

Fig. 10. Effect of horizontal flip and vertical flip on Hformer to fringe order prediction. (a), (b), (c), (d) and (e) respectively show the predicted fringe order by Hformer with Base, with vertical flipping, horizontal flipping, both vertical and horizontal flipping and the ground truth of fringe order.

## 4. Conclusion

In this work, we introduce the Transformer based on the self-attention mechanism into deep learning phase unwrapping and propose the Hformer model for fringe order prediction. The Hformer is designed to have a hybrid CNN-Transformer architecture based on CAT block in order to combine the advantages of CNN and Transformer. The encoder and decoder with cross attention are designed for the phase unwrapping problem. Experimental results demonstrate that our proposed Hformer performs better than U-Net and DCNN in terms of mIoU, MSE and MAE and in visual quality. Further, the ablation study on the proposed model was conducted to verify the effectiveness of the components in Hformer. We also verify that the improved FPN and strategies of augmentation using both horizontal flip and vertical flip can improve the predicted fringe order accuracy in ablation study.

It is demonstrated that, the Hformer is effective in fringe order prediction, as an alternative deep learning method to the existing CNN methods. Further works include the improvement of the proposed model to two frame wrapped phases in stereo based deep learning phase unwrapping or more frames in temporal based deep learning phase unwrapping, and the improvement of the proposed model in more complexed fringe projection 3D measurement conditions such as shadows, highlight, and projector distortion.

**Disclosures.** The authors declare no conflicts of interest.

**Data availability** Datasets used for training, validation and testing in this paper are available in Ref. [13]. Codes underlying this work may be obtained from the corresponding author upon reasonable request.